\documentclass[%
twocolumn
]{mpi2015-cscpreprint}

\usepackage{graphicx}      % include this line if your document contains figures

\usepackage[american]{babel}

\usepackage{todonotes}

\usepackage{geometry}
\usepackage{graphicx} % for pdf, bitmapped graphics files
\usepackage{float}
\usepackage{color}
\usepackage{url}
\usepackage{amsmath}
\usepackage{amsfonts}
\usepackage{amssymb}
\usepackage{scalefnt}

\usepackage[linesnumbered, ruled, vlined]{algorithm2e}

\newcommand{\bA}{{\mathbf A}}
\newcommand{\bB}{{\mathbf B}}
\newcommand{\bC}{{\mathbf C}}
\newcommand{\bD}{{\mathbf D}}
\newcommand{\bE}{{\mathbf E}}

\newcommand{\bH}{{\mathbf H}}
\newcommand{\bI}{{\mathbf I}}

\newcommand{\bS}{{\mathbf S}}

\newcommand{\bX}{{\mathbf X}}
\newcommand{\bY}{{\mathbf Y}}

\newcommand{\bh}{{\mathbf h}}

\newcommand{\bu}{{\mathbf u}}
\newcommand{\bv}{{\mathbf v}}

\newcommand{\bx}{{\mathbf x}}
\newcommand{\by}{{\mathbf y}}

\def\IR{{\mathbb R}}

\def\IL{{\mathbb L}}
\def\IV{{\mathbb V}}
\def\IW{{\mathbb W}}
\def\IH{{\mathcal H}}
\newcommand{\sIL}{{{{\mathbb L}_s}}}
\newcommand{\sIH}{{{{\mathcal H}_s}}}

\newtheorem{remark}{Remark}

\newtheorem{problem}{Problem}

\newcommand*{\victor}{\textcolor{black}}

\begin{document}
	
	\title{Data-driven approximation and reduction \\ from noisy data in matrix pencil frameworks} 

	\author[$\ast$]{Pauline Kergus}
\affil[$\ast$]{CNRS, LAPLACE, Toulouse, France.\authorcr
	\email{pauline.kergus@laplace.univ-tlse.fr}, \orcid{0000-0001-8601-0813}}
	
	\author[$\dagger$]{Ion Victor Gosea}
	\affil[$\dagger$]{Max Planck Institute for Dynamics of Complex Technical Systems, Magdeburg, Germany.\authorcr
		\email{gosea@mpi-magdeburg.mpg.de}, \orcid{0000-0003-3580-4116}}

	\shorttitle{On computing reduced-order models from noisy data}
	\shortauthor{P. Kergus and I. V. Gosea}
	\shortdate{}
	

%%%%%%%%%%%%%%%%%%%%%%%%%%%%%%%%%%%%%%%%%%%%%%%%%%%%%%%%%%%%%%%%%%%%%%%%%%%%%%%%
\abstract{ 
This work aims at tackling the problem of learning surrogate models from noisy time-domain data \victor{by means of} matrix pencil-based techniques, namely the Hankel and Loewner frameworks. A data-driven approach to obtain reduced-order state-space models from time-domain input-output measurements for \victor{linear time-invariant} (LTI) systems is proposed. \victor{This is accomplished by combining the aforementioned model order reduction (MOR)} techniques with the \victor{signal matrix model} (SMM) approach. The proposed method is illustrated by a numerical example consisting of a building model.}  
% The proposed technique allows to incorporate previously proposed solutions related to the solutions to improve the results.

\maketitle

%%%%%%%%%%%%%%%%%%%%%%%%%%%%%%%%%%%%%%%%%%%%%%%%%%%%%%%%%%%%%%%%%%%%%%%%%%%%%%%%

\section{Introduction}
\label{sec:introduction}

Numerous complex dynamical systems used in practical applications cannot be accurately described by physical models that are simple enough to be simulated or to be used for control design purposes. Model order reduction techniques then play a crucial role in obtaining a suitable complexity-accuracy trade-off. As recalled in \cite{ACA05}, MOR is usually based on full knowledge of a complex and high-fidelity system description, derived from physics laws. However, the increasing availability of data and the rise of data-driven applications require the incorporation of measurements when modeling or controlling a system. To that extent, data-driven reduction techniques, such as the Loewner Framework (LF) \cite{mayo2007framework}, Vector Fitting (VF) \cite{gustavsen1999rational} or Adapative Antoulas Anderson (AAA) \cite{nakatsukasa2018aaa}, are particularly appealing.

This paper focuses on the LF, which was mostly applied (with some exceptions) for noise-free data, obtained by simulating a high-fidelity model of the dynamical system under investigation. Indeed, as pointed out in \cite{lefteriu2010modeling}, LF is quite sensitive to noisy (perturbed) data. Physical modes of the system may only be included in the model at the expense of overmodeling, which generally leads to high variances and overfitting. As a result, noisy data complicates the selection of the system's order and may lead to high approximation errors. To tackle this issue, in \cite{lefteriu2010modeling} the poles were selected according to their residue norm to make the Loewner framework more robust to noise. This approach has also been used in the context of data-driven control based on the LF in \cite{kergus2018data}. In \cite{ionita2013lagrange}, it is suggested that the choice of the frequencies (interpolation points), as well as the partition of the corresponding data points, impacts the robustness with respect to noise. This idea was also explored in \cite{gosea2021data} and \cite{palitta2021efficient}, where different partitioning were studied for various numerical experiments. In \cite{embree2019pseudospectra}, the influence of the location and partition of the data points was studied through the pseudospectrum of the Loewner pencil. In \cite{drmavc2019learning}, it was shown that for Gaussian noise, the resulting Loewner model error grows at most linearly with the standard deviation of noise. In \cite{sahouli2018iterative}, the LF procedure was iterated, using data extracted from the model obtained at the previous iterations, in order to make the singular value decomposition of the Loewner pencil more accurate. In \cite{kabir2016loewner}, a Loewner model that is accurate with respect to noisy frequency-domain data was obtained, and then corrected through iterative least-square approximations.

This work primarily aims at proposing a way to obtain reduced-order models (ROMs) through matrix pencils techniques (LF and HF) that is more robust to noisy data. The objective is hence to enable the use of such techniques to obtain a ROM from measurements. In what we propose, the order is a tunable parameter, without considering available access to a high fidelity representation. To that extent, this work is at the crossroads of MOR and system identification (SI). The proposed method is based on matrix pencils approaches (LF and HF). The HF is rooted in realization theory \cite{schutter2000minimal} since it constructs a minimal LTI realization from Markov parameters, i.e. impulse response of discrete-time systems. Therefore, HF can be seen as a time-domain counterpart of LF \cite{ionita2012matrix}. In practice, the impulse response often has to be estimated from available input-output data. This is usually done through least-squares-based linear regression. In this work, we propose strategies for making this approach more robust to noisy data by using the SMM method introduced in \cite{yin2020maximum} (which allows estimating the impulse response of a system from noisy data).

To sum up, the proposed approach brings together aspects from MOR, realization theory and SI in an unified framework, which constitutes the main contribution of this work. Time-domain data, consisting of noisy input-output measurements, is used to estimate the finite impulse response of the system as in \cite{yin2020maximum}, which constitutes a non parametric characterization of the underlying LTI system. The finite impulse response is then used to obtain a reduced-order, explicit model through the HF/LF.  
% \pauline{should we say more?}
% \victor{I think that it is ok...very nice introduction!}

The rest of the paper is organized as follows. In Section \ref{sec:preliminaries}, the problem under investigation is formulated. Here, the Loewner and Hankel frameworks, which constitute the basis of this work, are recalled. The proposed approach is then introduced in Section \ref{sec:contribution} and a detailed specification on tuning its hyper-parameters is also provided. This method is then illustrated by a numerical example in Section \ref{sec:example}, which is the Los Angeles Hospital building benchmark from the COMP$l_eib$ library \cite{leibfritz2004compleib}, described by a 48th-order state-space model. Finally, the conclusion and outlook are discussed in Section \ref{sec:conclusion}.

\section{Preliminaries}
\label{sec:preliminaries}

\subsection{Problem formulation}

We consider an LTI discrete-time system with $n_u$ inputs, $n_y$ outputs and of order $n_x$, characterized by the following state-space realization:
% \begin{equation}\label{eq:linearSysDisc}
% \bH:\left\{\begin{aligned}
% \bE\bx_{t+1}&=\bA\bx_{t}+\bB u_{t}\\
% y_{t}&=\bC\bx_{t}+\bD u_{t}
% \end{aligned}\right. ,
% \end{equation}
\begin{equation}\label{eq:linearSysDisc}
\bH:\left\{\begin{aligned}
\bx_{t+1}&=\bA\bx_{t}+\bB u_{t}\\
y_{t}&=\bC\bx_{t}+\bD u_{t}
\end{aligned}\right. ,
\end{equation}
with $\bx \in \IR^{n_x}$ the state vector, $\bu \in \IR^{n_u}$ the input vector, $\by \in \IR^{n_y}$ the output vector, 
% $\bE \in \IR^{n_x \times n_x}$, 
$\bA \in \IR^{n_x \times n_x}$, $\bB \in \IR^{n_x \times n_u}$, $\bC \in \IR^{n_y \times n_x}$ and $\bD\in \IR^{n_y \times n_u}$. The value of a vector $v$ at the time step $t$ is denoted $\bv_t$. 
% \victor{It is considered that the matrix $\bE$ in (\ref{eq:linearSysDisc}) is invertible.}

The transfer function of (\ref{eq:linearSysDisc}) is given by:
% \begin{equation}
%     \bH(z) = \bD+ \bC (z \bE - \bA)^{-1} \bB.
% \label{eq:TransferFunction}
% \end{equation}
\begin{equation}
\bH(z) = \bD+ \bC (z\bI - \bA)^{-1} \bB.
\label{eq:TransferFunction}
\end{equation}

% Now, since the matrix $\bE$ is invertible, in can be incorporated into the other two system matrices in the difference equation from (\ref{eq:linearSysDisc}), as
% $\check{\bA} = \bE^{-1} \bA$ and $\check{\bB} = \bE^{-1} \bB$. Hence, we will have that $\check{\bE} = \bI$ (standard format), while the second equation in (\ref{eq:linearSysDisc}) does not change. Then, the transfer function of the transformed system is not changed either, as shown:
% \begin{align}\label{eq:TransferFunction2}
% \begin{split}
%     \check{\bH}(z) &= \bD+ \bC (z \check{\bE} - \check{\bA})^{-1} \check{\bB}\\
%     &= \bD+ \bC (z \bI - \bE^{-1} \bA)^{-1} \bE^{-1} \bB =     \bH(z).
%     \end{split} 
% \end{align}

% \victor{We will interchange between the descriptor and the standard format throughout the paper, since the descriptor format is needed in the LF and HF frameworks.}

The Loewner Framework (LF) \cite{mayo2007framework}, recalled in here in Section \ref{subsec:LF}, can identify the underlying system from noise-free frequency-domain samples $H(\imath \omega_i)$ in \eqref{eq:TransferFunction}. The Hankel Framework (HF) \cite{schutter2000minimal}, summarized here in Section \ref{subsec:HF}, relies on the impulse response $\left\{ \bh_k \right\}$ that connects the input and output samples as follows:
\begin{equation}\label{eq:impulseCoeffs}
y_{t}=\sum_{k=-\infty}^{\infty} \bh_{k} u_{t-k}.
\end{equation} 
The HF could be considered as a time-domain counterpart of the LF, as the frequency-domain representation \eqref{eq:TransferFunction} and the time-domain one \eqref{eq:linearSysDisc} are connected: the impulse response coefficients, also known as Markov parameters, are defined as follows:
% \victor{shall we present them like this or not?!}
% \begin{equation}
%     \bh_k = \begin{cases}
%     \bD, \ \ \text{if} \ \ k = 0, \\
%     \bC \left(\bE^{-1}\bA\right)^{k-1} \left(\bE^{-1}\bB\right), \ \ \text{if} \ \ k > 0,
%     \end{cases}
%     \label{eq:MarkovParameters}
% \end{equation}
\begin{equation}
\bh_k = \begin{cases}
\bD, \ \ \text{if} \ \ k = 0, \\
\bC \bA^{k-1}\bB, \ \ \text{if} \ \ k > 0,
\end{cases}
\label{eq:MarkovParameters}
\end{equation}
and they can be used to rewrite the transfer function \eqref{eq:TransferFunction} as an Infinite Impulse Response (IIR) filter:
\begin{equation}
\bH(z) =  \sum_{k=0}^\infty \bh_k z^{-k}.
\label{eq:IIRfilter}
\end{equation}
These matrix pencils techniques (LF and HF) also allow to reduce the order of the obtained models in a straightforward manner.

\begin{remark}[Descriptor/state-space forms]
	Both the LF and HF obtain a descriptor model. 
	\begin{equation}\label{eq:dss}
	\hat{\bH}:\left\{\begin{aligned}
	\hat{\bE}\hat{\bx}_{t+1}&=\hat{\bA}\hat{\bx}_{t}+\hat{\bB} u_{t}\\
	y_{t}&=\hat{\bC}\hat{\bx}_{t}+\hat{\bD} u_{t}
	\end{aligned}\right. ,
	\end{equation}
	In practice, as the considered system \eqref{eq:linearSysDisc} is causal, and due to the reduction process, the $\hat{\bE}$ matrix in \eqref{eq:dss} is full rank and therefore invertible. It is then possible to rewrite (\ref{eq:dss}) in a standard state-space form as in (\ref{eq:linearSysDisc}), i.e.:
	\begin{equation}\label{eq:Loewner_model}
	\hat{\bH}:\left\{\begin{aligned}
	\hat{\bx}_{t+1}&=\hat{\bE}^{-1}\hat{\bA}\hat{\bx}_{t}+\hat{\bE}^{-1}\hat{\bB} u_{t}\\
	y_{t}&=\hat{\bC}\hat{\bx}_{t}
	\end{aligned}\right. ,
	\end{equation}
	The same considerations hold for HF.
\end{remark}

While the LF and HF techniques have proven to be fairly successful when applied to MOR of given (large-scale) complex systems, they are indeed known to be quite sensitive to noisy data \cite{lefteriu2010modeling}. The problem under consideration that is tackled in this paper is formulated below: 

\begin{problem}
	\textit{Given noisy data, how to obtain a linear \\ reduced-order approximation of the underlying dynamical system through the LF or the HF frameworks?}
\end{problem}

In a sense, by using noisy data through these techniques, we more generally aim at bridging the gap between MOR, in which the underlying system is known but of complex or large-scale structure, and SI, which aims at building models from (noisy) measurements. As the proposed approach relies on the LF and HF frameworks, we will briefly summarize them in the two next subsections.

% \begin{remark}[Stability assumption]
% \pauline{I thought stability was required for \eqref{eq:impulseCoeffs} and \eqref{eq:IIRfilter}, to hold but it is not necessary. Why did we put that assumption already? Why is it necessary in our work to have a convergent Markov series?}

% \victor{Good question! I was thinking only through a numerical point of view...might cause numerical issues! But yes, I agree, does not have to be convergent!}

% \pauline{In case the system indeed has to be stable}: As the proposed approach relies on time-domain input-output measurements, collecting such data implies that the system is either stable or stabilized.
% \end{remark}

\begin{remark}[Discrete vs continuous] For continuous-time LTI systems, the relation between the Markov parameters $\bh$ and the impulse
	response $y_h$ is given by \cite{schutter2000minimal}:
	\begin{equation*}
	\bh_k = \left. \frac{d^{k-1}y_h(t)}{dt^{k-1}}\right|_{t=0} .
	\end{equation*}
	In practice, collecting input-output measurements implies that a continuous-time system appears as a sampled-data system, which is inherently discrete. Therefore, the rest of this work focuses on discrete-time systems. However, it should be noted that when working with the LF in the second part of the proposed approach, it is possible to obtain a continuous-time model as explained in \cite{vuillemin2021hybrid}.
\end{remark}

\begin{remark}[Time-domain LF]
	Another counterpart of LF in the time domain was proposed in \cite{peherstorfer2017data}: based on noise-free time-domain data $\{u_{k},y_{k}\}_k$ and on the knowledge of a high-fidelity model, frequency-domain data is inferred to use the LF. In comparison, the present work proposes to obtain a non-parametric characterization of the system by estimating its Markov parameters, from which frequency-domain data can be inferred to be used in the LF. Contrary to \cite{peherstorfer2017data}, the proposed approach does not require any description of the system and is more robust to noise.
\end{remark}

\subsection{The Loewner framework}
\label{subsec:LF}

Here, we briefly review the LF approach; for a more involved analysis, we refer the reader to the tutorial paper for LTI systems in \cite{ALI17}. The LF is based on frequency-domain measurements $\left\{\bH(z_k)\right\}_{k=1}^N$ corresponding to the transfer function \eqref{eq:TransferFunction}.

LF finds a state-space model $\hat{\bH}$ such that the following interpolation conditions are (approximately) fulfilled:
\begin{equation} \label{interp_cond}
\hat{\bH}(z_k)=\bH(z_k) ~\forall k=1 \dots N.
\end{equation}
The available data is partitioned into two disjoint subsets, $\left\{\bH(z_i)\right\}_{i=1}^\frac{N}{2}$ and $\left\{\bH(z_j)\right\}_{j=1}^\frac{N}{2}$. The Loewner pencil $\left(\IL,\sIL\right)$ is defined as follows
\begin{equation} \label{Loew_mat}
\IL_{(i,j)}=\frac{\bH(z_i)-\bH(z_j)}{z_i-z_j}, \ \sIL_{(i,j)}=
\frac{z_i\bH(z_i) -z_j \bH(z_j)}{z_i-z_j},
\end{equation}
while the data vectors $\IV, \IW^T \in \IR^k$ are introduced as
\begin{equation} \label{VW_vec}
\IV_{(i)}= \bH(z_i), \ \  \IW_{(j)} = \bH(z_j),~\text{for}~i,j=1,\ldots, \frac{N}{2}.
\end{equation}
By assuming that the data is not redundant, a minimal realization is then given by:
\begin{align*}
\hat{\bE}=-\IL,~~ \hat{\bA}=-\sIL,~~ \hat{\bB}=\IV,~~ \hat{\bC}=\IW.
\end{align*}
This typically means that no compression or reduction is required to identify the original model; more precisely, when $N = 2n_x$.

However, in practical applications, the Loewner pencil $(\sIL,\,\IL)$ is often singular (large quantities of data are processed) and can be hence challenging to use due to numerical issues. Hence, a ROM needs to be computed (with a corresponding Loewner pencil that is regular). In such cases, a  singular value decomposition (SVD) of the Loewner matrices is typically performed in order to determine a suitable truncation index $r$ and the corresponding projection matrices denoted with $\bX_r$ and $\bY_r$. The typical choice for computing these matrices is given below (as in eq. (8.38) from \cite{ALI17}):
\begin{equation}
\begin{bmatrix} \IL & \sIL \end{bmatrix} =\bX_1 \bS_1 \bY_1^*, \ \ \begin{bmatrix} \IL \\ \sIL \end{bmatrix} =\bX_2 \bS_2 \bY_2^*.
\end{equation}
Then, $\bX_r$ is chosen as the first $r$ columns of $\bX_1$, while $\bY_r$ as the first $r$ columns of $\bY_2$. To avoid enforcing polynomial terms such as a constant D-term (as it is the case of this work), the projection matrices are computed solely based on the SVD of the Loewner matrix $\IL$:
\begin{equation}
\label{eq:svd_Loewner}
~\IL =\bX \bS \bY^* \approx \bX_r\bS_r \bY_r^*.
\end{equation}	
Then, the reduced-order Loewner model of dimension $r$ is given by the following matrices:
\begin{equation}\label{Loew_red_lin}
\hat{\bE} = -\bX_r^*\IL \bY_r, \ \  \hat{\bA} = -\bX_r^*\sIL \bY_r, \ \
\hat{\bB} = \bX_r^*\IV, \ \  \hat{\bC} = \IW \bY_r.
\end{equation}

\begin{remark}[Data partitioning] How to effectively separate the available data into two subsets still remains an open question. It is shown in \cite{ionita2013lagrange} that this partition impacts the robustness to noise. In \cite{karachalios2021loewner,gosea2021data}, two different partitioning were numerically analyzed: 
	
	%\begin{table*}
	% \vspace{-\baselineskip}
	\begin{figure}[ht]
		% \small
		\begin{itemize}
			\item ``\textit{alternate}'' (the most recurrent way of separating data):
			\begin{equation}
			\left\{z_k\right\}_{k=1}^N=\left\{ \color{blue} z_1 , \color{red} z_2 , \color{black} \dots \color{blue} z_{N-1}, \color{red}z_N  \color{black} \right\}.
			\label{eq:alternateLF}
			\end{equation}
			\item ``\textit{half-half}'' (an intuitive way of separating data):
			\begin{equation}
			\left\{z_k\right\}_{k=1}^N=\left\{ \color{blue} z_1 , \dots, z_{N/2}, \color{red} z_{N/2+1}, \dots, z_N \color{black}. \right\}
			\label{eq:halfhalfLF}
			\end{equation}
		\end{itemize} 
		%  \vspace{.25\baselineskip}
		%  \hrule
		\caption{Splitting schemes commonly used in the LF}
	\end{figure}
	%\end{table*}

	%\victor{Add figure environment}

	As previously reported in \cite{gosea2021data}, the effect of half-half partitioning is that the decay of the singular values of the Loewner matrix is clearer (more revealing) than for the alternate splitting (when dealing with noisy frequency-domain data). As a result, half-half LF seems to ease the order selection and hence avoids overfitting due to noise. Both types of partitioning are used jointly in this work, as explained in the next section.
	\label{remark:data_partitioning}
\end{remark}

\subsection{The Hankel framework}
\label{subsec:HF}

While the LF interpolates the frequency response, the HF provides a model that interpolates the impulse response, similarly to the Ho-Kalman algorithm \cite{HoKal66} or Silverman realization \cite{Silv71}. Given the truncated impulse response $\bh = [\bh_0,\bh_1,\cdots,\bh_{N-1}]$, the resulting Hankel model is given in descriptor form, see \eqref{eq:Loewner_model}, by the following matrices:
\begin{equation}
\label{eq:Hankel_model}
\begin{array}{c}
\hat{\bE}=\IH,~~ \hat{\bA}=\sIH,~~ \\ 
\hat{\bC} = \big[\,\bh_1,~ \bh_2, \cdots,~ \bh_{N/2}\,\big],~~ \hat{\bB} = \hat{\bC}^T,~~ \hat{\bD} = \bh_0.
\end{array}
\end{equation}
with the Hankel pencil $\left(\IH,\sIH\right)$ defined as follows
\vspace{-2mm}
\begin{align}\label{eq:Hankel_pencil}
\small
\begin{split}
\IH &= \!
\left[\begin{array}{cccc}
\bh_1& \bh_2 & \cdots & \bh_\frac{N}{2}\\
\bh_2& \bh_3 & \cdots & \bh_{\frac{N}{2}+1} \\
\vdots & \vdots & \ddots & \vdots \\
\bh_\frac{N}{2} \hspace{-1mm}& \bh_{n+1}\hspace{-1mm} & \cdots & \bh_{N-1}
\end{array}\right]\!, \\
\sIH &= \!
\left[\begin{array}{cccc}
\bh_2& \bh_3 & \cdots & \bh_{\frac{N}{2}+1}\\
\bh_3& \bh_4 & \cdots & \bh_{\frac{N}{2}+2} \\
\vdots & \vdots & \ddots & \vdots \\
\bh_{\frac{N}{2}+1} \hspace{-1mm}& \bh_{\frac{N}{2}+2}\hspace{-1mm} & \cdots & \bh_{N}
\end{array}\right]\!.
\end{split}
\end{align}
As in the LF, the dimension of the Hankel model \eqref{eq:Hankel_model} can be reduced by means of projection, using orthogonal matrices computed by means of applying an \textbf{SVD} for the Hankel matrix $\IH$. In this case, we enforce approximation, i.e. by fitting a model which approximately explains the data. Additional insights on the HF were given in \cite{ionita2012matrix}.

\section{From noisy data to reduced-order models}
\label{sec:contribution}

\subsection{Overview of the proposed approach}

To the best of our knowledge, most of the attempts to make the LF and HF matrix pencils identification techniques more robust to noisy data have consisted in changing the way the model is obtained \cite{lefteriu2010modeling}, \cite{kabir2016loewner}, \cite{sahouli2018iterative}. In this work, it is proposed to preprocess the noisy data instead. 

% An overview of the proposed approach is given in Algorithm \ref{}.
First, an estimation of the truncated impulse response $\{\tilde{\bh}_k\}_{k=0}^{N-1}$ of the system is obtained from the available noisy measurements through the SMM approach, as proposed in \cite{yin2020maximum}. This estimation forms a non-parametric model of the system, which is then parameterized and reduced through the HF or the LF.

While the estimated values $\{\tilde{\bh}_k\}_{k=0}^{N-1}$ can be used directly in the HF, another possibility consists in applying a fast Fourier transform to the impulse response to estimate frequency-response samples as follows:
\begin{equation}
\Tilde{\bH}_N(e^{\imath \omega_i})= \sum_{k=0}^{N-1} \bh_k e^{-\imath \omega_i k},~ \omega = \frac{2\pi i}{N}, ~i=0\dots N-1,
\label{eq:FFT}
\end{equation}
which is a truncated version of \eqref{eq:IIRfilter}. The frequency-domain data estimated from \eqref{eq:FFT} can then be used in the LF.

In what follows, the SMM approach from \cite{yin2020maximum} is recalled in Section \ref{subsec:h_estimation}. The tuning knobs of the proposed approach, that combines SMM and matrix pencils approaches, are then detailed in Section \ref{subsec:tuning_knobs}. A synthesized algorithm that brings these different aspects together is then provided in Section \ref{subsec:algorithm}.

\subsection{Impulse response estimation: the SMM approach}
\label{subsec:h_estimation}
Traditionally, Markov parameters $\bh_k$'s can be obtained from input-output measurements $\left\{u_k,y_k\right\}_{k=0}^{N_s}$ by solving a linear system of equations, as:
\begin{equation}\label{eq:LSestimate}
\small
\left[ \begin{array}{c} y_{N-1} \\ y_{N} \\ \vdots \\ y_{N_s} \\
\end{array} \right]
=
\left[ \begin{array}{cccc} 
u_{N-1} & u_{N-2} & \dots & u_0 \\ 
u_{N} & u_{N-1} & \dots & u_1 \\ 
\vdots & \vdots & & \vdots \\ 
u_{N_s} & u_{N_s-1} & \dots & u_{Ns-N+1}  \\
\end{array} \right]
\left[ \begin{array}{c} \bh_0 \\ \bh_1 \\ \vdots \\ \bh_{N-1} \\
\end{array} \right],
\end{equation}
which is rewritten $\mathbf{U}_{N_s,N}\bh=\mathbf{Y}_{N_s,N}$.

This procedure allows estimating $N < \frac{N_s}{2}$ Markov parameters. It is important to note that in the noise-free case, truncating the Markov series will impact the results. Indeed, solving \eqref{eq:LSestimate} consists in identifying a $N$-th order FIR filter from the available data, rather than obtaining the true value of the first $N$-th Markov parameters. Assuming that the truncation error of the FIR is negligible, the least-squares solution of \eqref{eq:LSestimate} is known to be the best unbiased estimator for independent and identically distributed (i.i.d.) Gaussian output noise \cite{ljung1987theory}. In practice, a very long impulse response sequence may be needed to reach a negligible truncation error, even for a low-order system. The least squares (LS) approach, which solves \eqref{eq:LSestimate} in order to estimate the impulse response, then requires a significant amount of data and becomes computationally expensive, hence unfeasible. 

In \cite{markovsky2005data}, a data-driven simulation approach, based on Willems' fundamental lemma, was proposed when noise-free input-output data are available. It allows to estimate the impulse response even when the truncation error is not negligible. The following assumptions are enforced:

\begin{enumerate}
	\item The LTI system under consideration, i.e., $\bH$ in \eqref{eq:linearSysDisc} is finite-dimensional and controllable;
	\item The input $\left\{u_k\right\}_{k=0}^{N_s}$ is persistently exciting of order $L=N+n_x$, meaning that the Hankel matrix 
	\begin{equation}
	\label{eq:persistency}
	\mathcal{U}=\left[\begin{array}{cccc} 
	u_0 & u_{1} & \dots & u_M \\ 
	u_{1} & u_2 & \dots & u_{M+1} \\ 
	\vdots & \vdots & & \vdots \\ 
	u_{L-1} & u_{L} & \dots & u_{N_s}  \\
	\end{array}\right]\in \IR^{Ln_u\times M},
	\end{equation}
	with $M= N_s-L + 1$, has full row rank \cite{willems2005note}. 
\end{enumerate}
The parameter $L$ is to be chosen by the user and corresponds to the length of the trajectory to be estimated in this data-driven framework. More precisely, in the present work it denotes the number of Markov parameters to be estimated.

Under these assumptions, the output trajectory of the system for an input $\bu\in\IR^{Nn_u}$, starting from initial conditions uniquely determined by the past input trajectory $\bu_{ini}\in \IR^{L_0n_u}$ and $\by_{ini}\in \IR^{L_0n_y}$ for $L_0\geq n_x$, is $\by=Y_fg$. Here, $g\in \IR^{M'}$ is the solution of the linear system of equations:
\begin{equation}
\label{eq:implicit_model}
\left[\begin{array}{c} 
\bu_{ini} \\ 
\by_{ini} \\
\bu \\
\end{array}\right]=\left[\begin{array}{c} 
U_p \\ 
Y_p \\ 
U_f \\
\end{array}\right]g,
\end{equation}
where $U_p$, $U_f$, $Y_p$ and $Y_f$ are matrices computed by using the available data as follows: 
\begin{align}\label{eq:Hankel_pencil2}
\begin{split}
U_p &= \!
\left[\begin{array}{cccc} 
u_0 & u_{1} & \dots & u_{M'-1} \\ 
\vdots & \vdots & \ddots& \vdots \\ 
u_{L_0-1} & u_{L_0} & \dots & u_{M'+L_0-2}  \\
\end{array}\right]\! \in \mathbb{R}^{L_0n_u\times M'}, \\
U_f &= \!
\left[\begin{array}{cccc}
u_{L_0} & u_{L_0+1} & \dots & u_{M'+L_0-1}  \\
\vdots & \vdots & \ddots & \vdots \\ 
u_{L'-1} & u_{L'} & \dots & u_{N_s-1}  \\
\end{array}\right]\! \in \mathbb{R}^{Nn_u\times M'},
\end{split}
\end{align}
with $M'= N_s-L'+ 1$ and $L'= N + L_0$, and similarly for $Y_p$ and $Y_f$.

In order to handle the case for which only noisy input-output measurements are available, the SMM approach in \cite{yin2020maximum} builds on \cite{markovsky2005data} and represents a maximum likelihood framework to obtain a statistically optimal implicit model. Additive i.i.d Gaussian output noise is considered:
\begin{equation}
\tilde{\by} = \by + w, ~w\sim \mathcal{N}(0,\sigma^2\mathbb{I}).
\label{eq:output_noise}
\end{equation}
As in \cite{yin2020maximum}, the SMM approach is used to estimate the impulse response with $\bu_{ini}=0$, $\by_{ini}=0$ and $\bu=[1 ~ 0 \dots 0]$. The estimate of the first $N$ Markov parameters denoted with $\hat{\bh}$ is explicitly given by $\hat{\bh} = Y_f g_h$, where
\begin{align}\label{eq:SMM_estimator}
\begin{split}
g_h = ~ &  \! (F^{-1}-F^{-1}U^T(UF^{-1}U^T)^{-1}UF^{-1})Y_p^T \by_{ini}
\! \\
& + F^{-1}U^T(UF^{-1}U^T)^{-1} \!
\left[\begin{array}{c}
\bu_{ini}\\
\bu \\
\end{array}\right]\!, \\
F = ~  & Y_p^TY_p+L'\sigma^2\mathbb{I}, \ \ \text{and} \ \
U = ~  \left[\begin{array}{c}
U_p\\
U_f \\
\end{array}\right].
\end{split}
\end{align}

The result is unbiased for an arbitrary length $N$, as long as the input is persistently exciting of rank $N+n_x$.

\subsection{Tuning the hyper-parameters}
\label{subsec:tuning_knobs}

The success of the proposed approach depends on carefully choosing some particular parameters. Some hints on choosing these are provided in the next subsection.

\subsubsection{a) Persistency of excitation:}

Persistency of excitation is the key assumption of Willems' fundamental lemma \cite{willems2005note} as it allows to characterize all possible trajectories of length $N$ from the available data. However, assuming that the input should be persistently exciting of order $N+n_x$ implies that the order of the underlying system is known. To overcome this issue when the system is unknown, the SMM approach introduces $L_0$ instead and the matrix $U$, defined in \eqref{eq:SMM_estimator}, that should be of full row rank. In the ideal case, the value $L_0 = n_x$ should be used in order to exploit the available data to the fullest extent. Nonetheless, the most important condition to be imposed is $L_0\geq n_x$ so that $\bu_{ini}$ and $\by_{ini}$ uniquely define the initial conditions.

In practice, when $n_x$ is unknown, a good choice for $L_0$ can be found by computing the cross-correlation $R_{yu}$ of the measured output and the input signal:
\begin{equation}
\label{eq:cross_correlation}
R_{yu}(\tau)= \sum_{k=0}^{} y_{k+\tau}u_{k},
\end{equation}
and $L_0$ is then chosen as the minimal positive lag such that:
\begin{equation}
\label{eq:choice_L0}
\forall \tau > L_0, ~ |R_{yu}(\tau)|\leq \epsilon.
\end{equation}
As the system is causal, the cross-correlation for negative lags is merely a numerical artifact and does not represent any real input-output relationship. For this reason, the threshold $\epsilon$ is fixed in this work as:
\begin{equation}
\label{eq:threshold}
\epsilon = (1+\alpha)\times\textnormal{max} \left\{ |R_{yu}(\tau)| \textnormal{ for } \tau<0 \right\},
\end{equation}
where the scalar $0 \leq \alpha \leq 1$ allows introducing an additional margin to avoid choosing to large of a value for $L_0$ (as illustrated in Section \ref{sec:example}).

\subsubsection{b) The choice of $N$ as the number of estimated Markov parameters:}

A necessary condition for the matrix $U$ to be of full row rank (the so-called persistency of excitation assumption) is $M'\geq L'n_u$, which gives an upper bound $N_{max}$ for the number of Markov parameters (denoted with $N$) that can be estimated when a number of $N_s$ input-output measurements are available:
\begin{equation}
N_{max} = \frac{N_s+1}{n_u+1}-L_0.
\label{eq:bound_N}
\end{equation}
On the other hand, as the Hankel and shifted Hankel matrices are of size $N$, then $N$ Markov parameters allow to obtain a model of order at most $N$ through HF. In addition, when using the LF, the more Markov parameters are used, the lower the truncation error between \eqref{eq:IIRfilter} and \eqref{eq:FFT} becomes. 

Consequently, after having chosen $L_0$ as previously explained, it is recommended to choose $N = N_{max}/2$ to enforce $M'\ll L'n_u$. Alternatively, one could decrease it if necessary until the matrix $U$ is of full row rank. 

% Note that if more Markov parameters are necessary, the SMM approach can be used in an iterative fashion to obtain more Markov parameters.

\subsubsection{c) Noise variance $\sigma^2$:}

The noise variance $\sigma^2$ is used in the SMM approach to estimate a non-parametric model of the system, see \eqref{eq:SMM_estimator}. In practice, this information might not be available. An approximation can be obtained through the LS approach \eqref{eq:LSestimate}. As recalled in \cite{niu1995simultaneous}, for zero-mean white noise, an unbiased estimate of the variance $\sigma^2$ is given by:
\begin{equation}
\label{eq:variance_LS}
\sigma^2 = \underset{N_s\rightarrow \infty}{\textnormal{lim}} \frac{\Vert \mathbf{U}_{N_s,N}\bh_{LS}-\mathbf{Y}_{N_s,N}\Vert_2^2}{N_s-N},
\end{equation}
where $\bh_{LS}$ is the estimate by the classical LS approach from \eqref{eq:LSestimate}, obtained by using $N_s$ input-output samples.An approximation $\hat{\sigma}^2$ is then chosen as follows:
\begin{equation}
\label{eq:variance_estimate}
\hat{\sigma}^2 = \frac{\Vert \mathbf{U}_{N_s,N}\bh_{LS}-\mathbf{Y}_{N_s,N}\Vert_2^2}{N_s-N}.
\end{equation}

\subsubsection{d) Order of the reduced-order model:}

The order of the reduced-order model is a tunable parameter for both HF and LF. An adequate value is supposed to be chosen based on a rank-revealing decomposition of the Hankel or Loewner matrices. As detailed in \cite{lefteriu2010modeling}, measurement noise complicates the choice of the reduced order $r$. In that case, it is possible to change the data partitioning in the LF in order to obtain a clearer SVD decay, as suggested in \cite{gosea2021data} and recalled in Remark \ref{remark:data_partitioning}. However, while half-half partitioning \eqref{eq:halfhalfLF} reveals the system's order in a clear way and is robust to noise, it leads to less accurate models. This is because the Loewner matrices tend to be  ill-conditioned for this choice. At the same time, the SVD resulting from alternate partitioning \eqref{eq:alternateLF} is more visibly affected by noise. Hence, this makes it challenging to choose the order. However, this choice typically leads Loewner pencils that are diagonally dominant. Therefore, for a fixed order, this approach will result in more accurate models. This behavior has been pointed out in \cite{ionita2013lagrange}, and more recently in \cite{palitta2021efficient} based on analyzing Cauchy matrices, which explicitly appear in the definition of Loewner matrices.

% \pauline{Can you check that it's okay?}

% \victor{Yes}

For such reasons, we propose here to combine both types of data-partitioning in the LF to benefit from their respective advantages. The Loewner matrix $\IL_{hh}$ based on half-half partitioning \eqref{eq:halfhalfLF} is first put together and its SVD is performed to select the truncation order $r$. Then, the Loewner pencil $\left(\IL_{alt},\sIL_{alt}\right)$ based on alternate partitioning \eqref{eq:alternateLF} is computed in order to obtain a model of order $r$ as in \eqref{Loew_red_lin}.

\subsection{Summary}
\label{subsec:algorithm}

Given noisy data $\{u_k,\tilde{y}_k\}, \ k=0 \ldots N_s-1$, the proposed approach consists in tuning some hyper-parameters as explained in Section \ref{subsec:tuning_knobs}, before using the SMM approach from \cite{yin2020maximum} as recalled in Section \ref{subsec:h_estimation}. The resulting estimated Markov parameters $\bh_{SMM}$, which constitute a non-parameterized model of the system, are then used in matrix pencil approaches, the HF (Algorithm \ref{SMM-HF}) or the LF (Algorithm \ref{SMM-LF}), allowing to obtain a linear reduced-order approximation $\left(\hat{\bE}_r,\hat{\bA}_r, \hat{\bB}_r, \hat{\bC}_r, \hat{\bD}_r \right)$ of the underlying dynamical system \eqref{eq:linearSysDisc}.

These two techniques are referred to as SMM-HF and SMM-LF respectively. The main advantage of SMM-LF is that it allows using different data-partitioning techniques in order to reveal the order of the system despite measurement noise. However, it requires to build an additional matrix $\IL_{hh}$ and the corresponding SVD. Finally, it should be noted that the frequency-domain data estimated from \eqref{eq:FFT} when using SMM-LF is affected by the truncation of the Markov series when the truncation error is not negligible, while the SMM-HF is not sensitive to it. For this reason, it might be more interesting to use the HF once the order $r$ has been determined from the SVD of $\IL_{hh}$, which combines SMM-HF and SMM-LF.

\begin{algorithm}
	\textbf{Inputs:} Input-output time-domain data $\{u_k,\tilde{y}_k\}, \ k=0 \ldots N_s-1$.
	\begin{enumerate}
		\item Step 1: Tuning the hyper-parameters
		\begin{enumerate}
			\item Compute the cross-correlation $R_{yu}$ \eqref{eq:cross_correlation}, the threshold $\epsilon$ \eqref{eq:threshold} and then $L_0$ \eqref{eq:choice_L0}.
			\item Choose $N=N_{max}/2$ based on \eqref{eq:bound_N} and decrease it until $U$ is full rank.
			\item Solve \eqref{eq:LSestimate} to obtain $\bh_{LS}$, then estimate the noise variance $\hat{\sigma}^2$ \eqref{eq:variance_estimate}.
		\end{enumerate}
		\item Step 2: Using $L_0$, $N$ and $\hat{\sigma}^2$, estimate the Markov parameters $\bh_{SMM}$ through SMM \eqref{eq:SMM_estimator} as in \cite{yin2020maximum}.
		\item Step 3: Apply the HF
		\begin{enumerate}
			\item Build the Hankel pencil $\left(\IH,\sIH\right)$ \eqref{eq:Hankel_pencil} based on $\bh_{SMM}$.
			\item Perform an SVD of $\IH$ to determine the order $r$ and compute projection matrices $\bX_r$ and $\bY_r$:
			$$\IH =\bX \bS \bY^* \approx \bX_r\bS_r \bY_r^*.$$
			\item Build the Hankel model \small $\hat{\bE}=X_r^*\IH Y_r$, $\hat{\bA}=X_r^*\sIH Y_r$, $\hat{\bC} = \big[\,\bh_1,~ \bh_2, \cdots,~ \bh_N\,\big]Y_r$, $\hat{\bB} = X_r^*\big[\,\bh_1,~ \bh_2, \cdots,~ \bh_N\,\big]^T$, $\hat{\bD} = \bh_0$.
		\end{enumerate}
	\end{enumerate}
	\caption{SMM-HF}
	\label{SMM-HF}
\end{algorithm}

\begin{algorithm}
	\textbf{Inputs:} Input-output time-domain data $\{u_k,\tilde{y}_k\}, \ k=0 \ldots N_s-1$.
	\begin{enumerate}
		\item Steps 1 and 2: same as in Algorithm \ref{SMM-HF}.
		\item Step 3: Apply the LF
		\begin{enumerate}
			\item 		\hspace{-3mm} Perform the FFT of $\bh_{SMM}$ to infer frequency-domain data \eqref{eq:FFT}.
			\item \hspace{-3mm} Build the Loewner matrix $\IL_{hh}$ based on half-half partitioning \eqref{eq:halfhalfLF}.
			\item \hspace{-3mm} Perform an SVD of $\IL_{hh}$ to appropriately determine the order $r$.
			\item \hspace{-3mm} Build the Loewner $\left(\IL_{alt},\sIL_{alt}\right)$ \eqref{Loew_mat} based on the alternate partitioning \eqref{eq:alternateLF},
			and on matrices $\bX_r$ and $\bY_r$ from \eqref{eq:svd_Loewner} of dimension $r$, chosen previously.
			\item \hspace{-3mm} Build the Loewner model \eqref{Loew_red_lin}.
		\end{enumerate}
	\end{enumerate}
	\caption{SMM-LF}
	\label{SMM-LF}
\end{algorithm}

\section{Numerical example}
\label{sec:example}

The proposed approach is illustrated on the Los Angeles Hospital building benchmark from the COMP$l_eib$ library \cite{leibfritz2004compleib}, described by a 48th-order state-space model. It was originally used in \cite{antoulas2000survey}. The system has one input and one output. It should be noted that the proposed approach is also applicable to multivariable systems.
% \pauline{Apparently the system is not controllable, but the A matrix is ill-conditionned so maybe the classic matlab check is not sufficient}

% \victor{I just checked the A matrix for the "build" model below:}

% \begin{center}
%     \url{http://slicot.org/20-site/126-benchmark-examples-for-model-reduction}
% \end{center}

% \victor{and I noticed that the condition number of the A matrix is $1.22 \cdot 10^4$ - so not that bad!}

% \begin{center}
%     \includegraphics[scale=0.22]{SV_P_Gramian.eps}
% \end{center} 

To collect data, the high-order model is simulated using a normally distributed random input signal. The sampling period is $T_s=15$ms and $N_s=1000$ output samples $y_k$ are collected. Additive output Gaussian noise of variance $\sigma^2=1\cdot 10^{-7}$ is then considered, as in \eqref{eq:output_noise}. 50 different noisy data sets are generated like this.

% Different noise levels are considered in the end of this section to highlight its impact on modeling errors, as detailed in \ref{subsec:noise_level}.

Algorithms \ref{SMM-HF} and \ref{SMM-LF} are derived hereafter. The first two steps are common to SMM-HF and SMM-LF and are detailed in Sections \ref{subsec:step1} and \ref{subsec:step2}. Obtaining a reduced and parameterized model through LF and HF is then detailed in Section \ref{subsec:step3}. 

\subsection{Step 1: Choice of the hyper-parameters}
\label{subsec:step1}

\begin{figure}[h]
	\centering
	\includegraphics[width=0.45\textwidth]{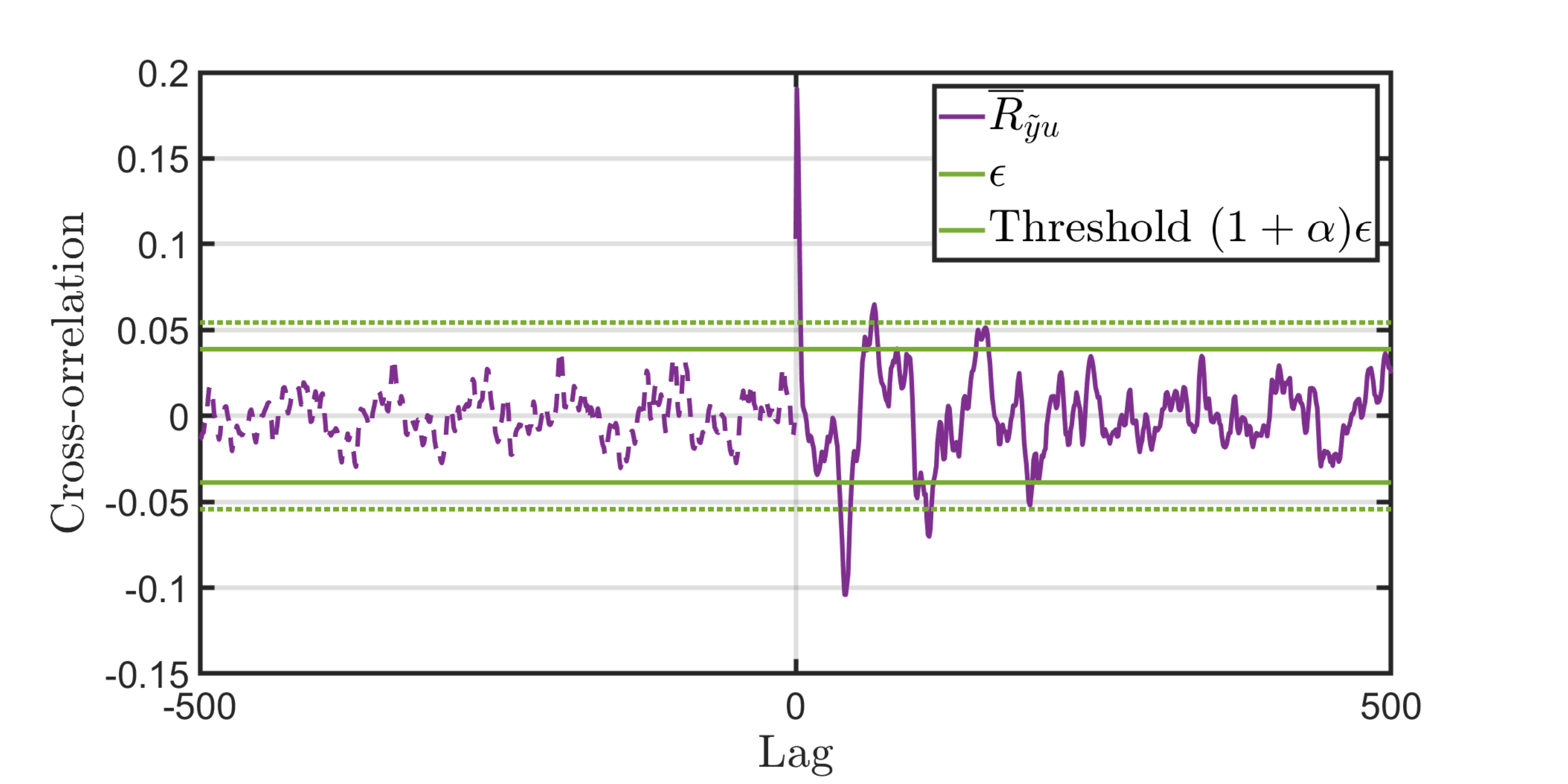}
	\vspace{-3mm}
	\caption{Choice of $L_0$ using the average cross-correlation $\overline{R}_{\tilde{y}u}$ ($L_0=66$).}
	\vspace{-3mm}
	\label{fig:choice_L0}
\end{figure}

The cross-correlation is computed for every noise realization and averaged. As represented on Figure \ref{fig:choice_L0}, the threshold value is chosen as in \eqref{eq:threshold} with $\alpha = 0.4$ and, according to \eqref{eq:choice_L0}, $L_0=66$ is taken, which slightly overestimates the order $n_x$ of the system. In the present case, $\alpha = 0.4$ is chosen to eliminate the cross-correlation coefficient that are very close to the threshold $\epsilon$.

The number of estimated Markov parameters is taken equal to $N=N_{max}/2=217$, and the corresponding matrix $U$ is full row rank, which means that the input $u$ is persistently exciting of order $L_0+N$.

The LS approach is applied and the resulting variance estimate is $\hat{\sigma}^2 = 1.27\cdot 10^{-7}$.

\subsection{Step 2: Impulse response estimation}
\label{subsec:step2}
Based on each noisy dataset $\left\{u_k,\tilde{y}_k\right\}_{k=0}^{N_s-1}$, the SMM approach is used to estimate the first $N$ Markov coefficients of the system, i.e. the first $N$ samples of its impulse response. As in \cite{yin2020maximum}, the fitting of the estimated impulse response $\hat{\bh}$ to the true system impulse response $\bh$ is defined by:
\begin{equation}
\label{eq:fit}
W=100\left(1-\sqrt{\frac{\sum_{i=1}^N(\bh_i-\hat{\bh}_i)^2}{\sum_{i=1}^N(\bh_i-\overline{\bh})^2}}\right) , 
\end{equation}
with $\overline{\bh}$ the average of the true Markov parameters $\bh$. The results correspond to the level of performance presented in \cite{yin2020maximum}: the SMM approach ($W=54.2\%$) outperforms the LS ($W=47\%$) one by obtaining a better median fit.

\subsection{Step 3: Model approximation and reduction}
\label{subsec:step3}

The estimated impulse responses, denoted $\bh_{LS}$ and $\bh_{SMM}$ for the LS and SMM approach respectively, obtained in Step 2, one fore each noisy data set, are now used to obtain a parameterized model of the system through LF and HF.

\subsubsection{a) Loewner framework:}
Frequency-domain data is inferred by performing a fast Fourier transform as in \eqref{eq:FFT} of the SMM estimated impulse response. For comparison purposes, frequency-domain data is also estimated as the ration between the cross power spectral density of $u$ and $y$, and the power spectral density of $u$, without taking noise into account. This last approach is referred to as \textit{noisy LF} in this paragraph.

\begin{figure}[h]
	\centering
	\includegraphics[width=0.45\textwidth]{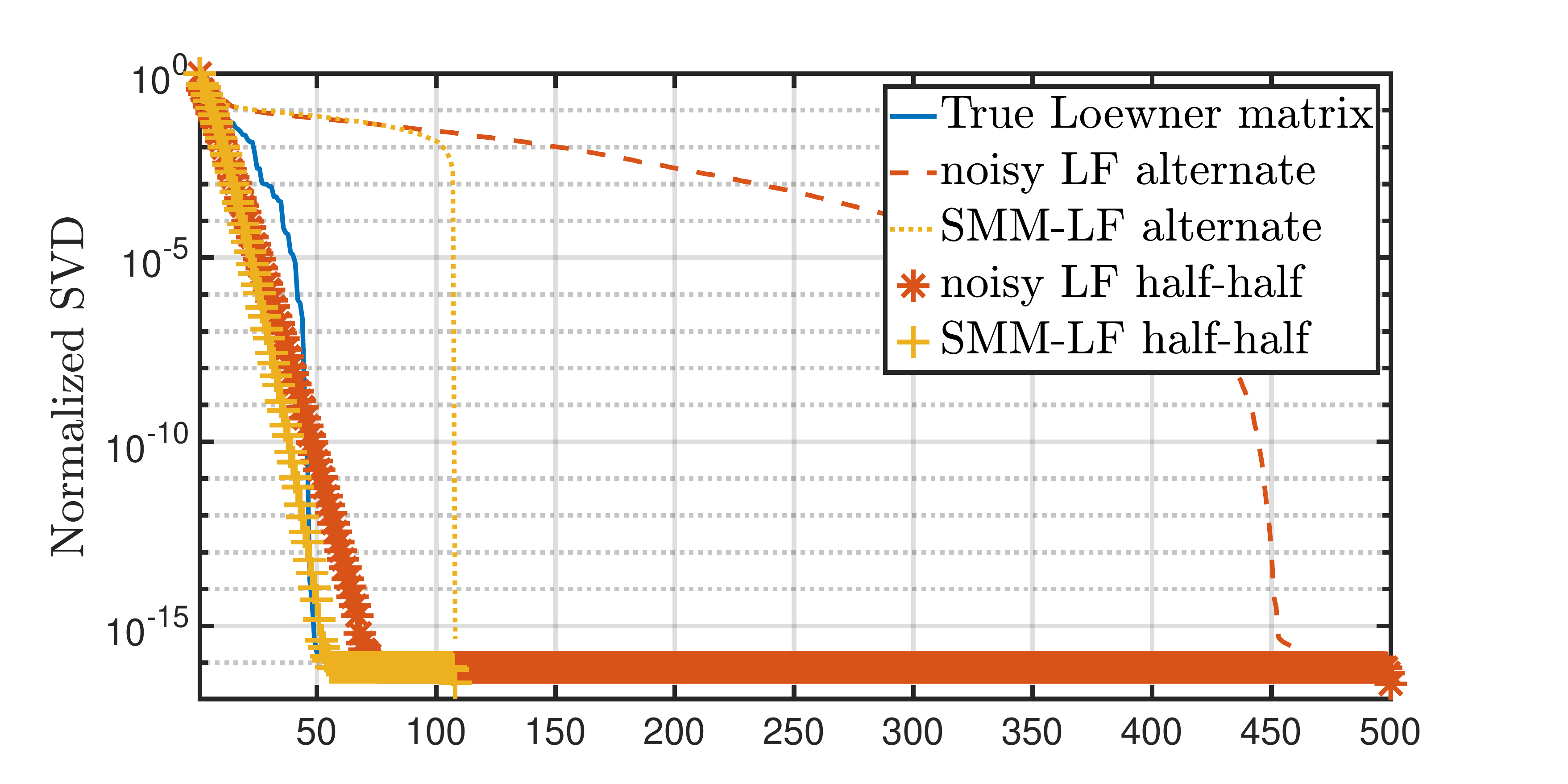}
	\caption{Normalized SVD of the Loewner matrices built with the frequency-domain data inferred from the SMM approach or directly from the noisy time-domain data (\textit{noisy LF}), and compared with the SVD decay of the Loewner matrix obtained with noise free frequency-domain data. Two types of data partitioning are used as in \cite{gosea2021data} to evaluate the order of the underlying system despite measurement noise.}
	\label{fig:SVD_Loewner}
\end{figure}

Once frequency-domain data is obtained, the Loewner pencil from \eqref{Loew_mat} is then built using the two different data partitioning techniques presented in \cite{gosea2021data} and recalled in \eqref{eq:alternateLF} and \eqref{eq:halfhalfLF}. A SVD is performed on the Loewner matrix $\IL$ to reveal the order of the underlying system. The average decay of the normalized singular values is visible on Figure \ref{fig:SVD_Loewner}: while alternate partitioning gives almost full-rank Loewner matrices with both the noisy LF and SMM-LF approaches, half-half partitioning leads to a Loewner matrix of order $48$ for the SMM-LF approach and $60$ for the noisy-LF approach (in average over the 50 noisy data sets). If allowing to approximate the order of the underlying system, half-half partitioning leads to less precise models, as highlighted in \cite{gosea2021data}. Descriptor models are then obtained as in \eqref{Loew_red_lin}, based on alternate partitioning \eqref{eq:alternateLF} as suggested in Algorithm \ref{SMM-LF}. The order is chosen as $r= n_x = 48$. 

\begin{figure}[h]
	\centering
	\includegraphics[width=0.45\textwidth]{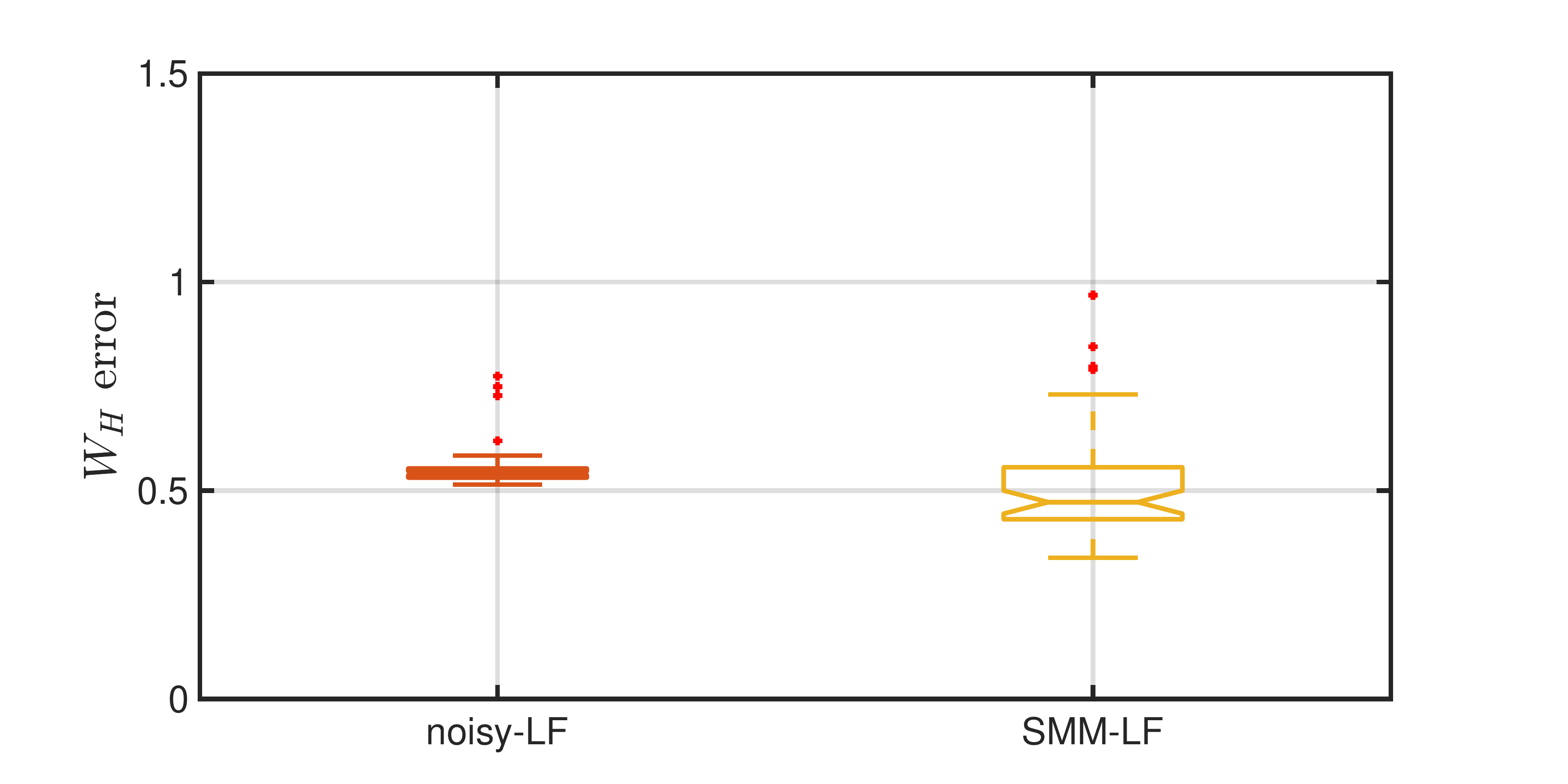}
	\vspace{-3mm}
	\caption{Normalized $\mathcal{H}_2$-error $W_H$ between the resulting frequency-responses and the one of the true system.}
	\label{fig:WH_SMM_LF}
	\vspace{-3mm}
\end{figure}

\begin{figure}[h]
	\centering
	\includegraphics[width=0.45\textwidth]{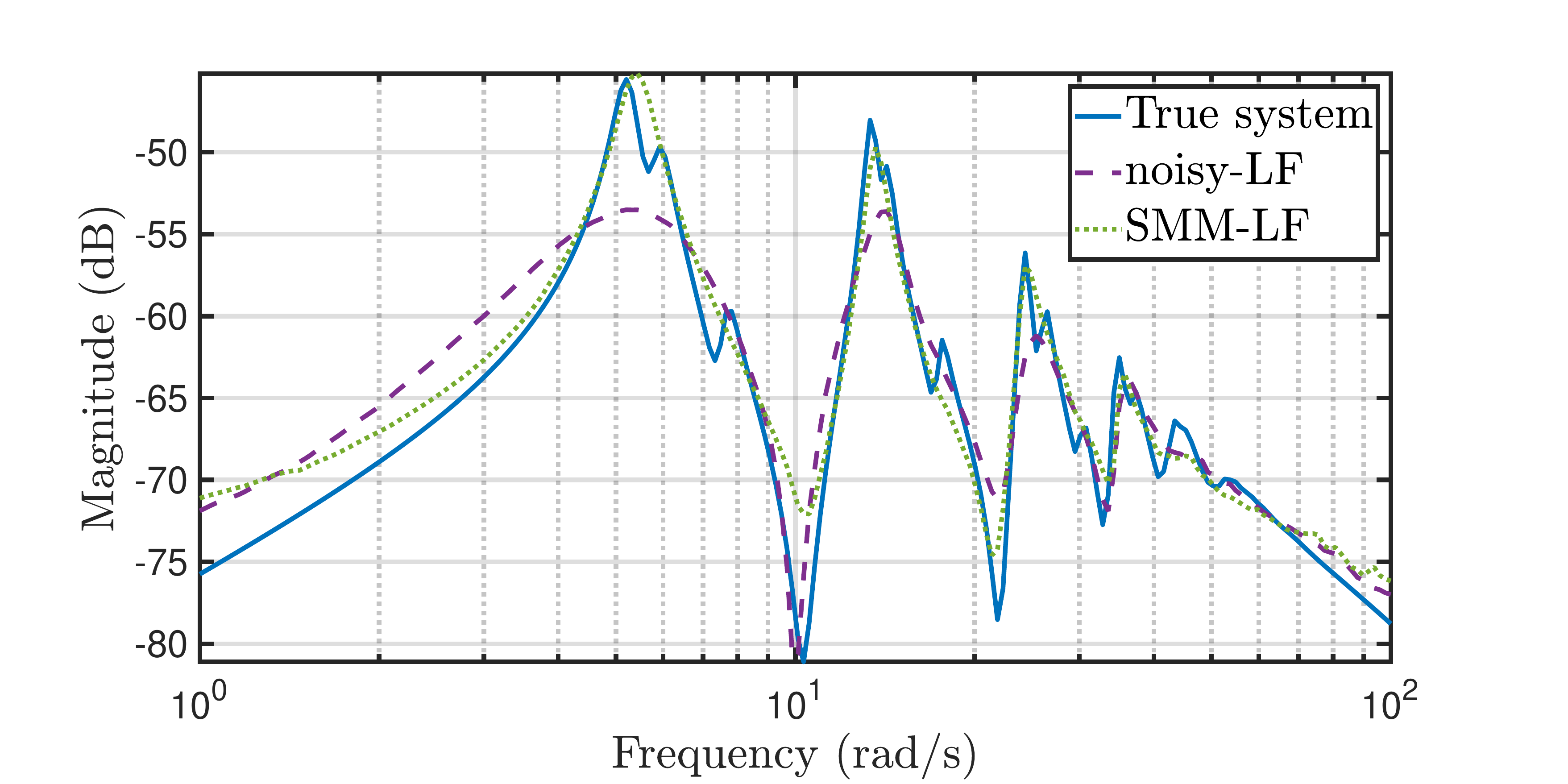}
	\vspace{-3mm}
	\caption{Average frequency-response obtained when applying the SMM-LF and noisy LF procedures.}
	\label{fig:LF_freqresp}
	\vspace{-3mm}
\end{figure}

\begin{figure}[h]
	\centering
	\includegraphics[width=0.45\textwidth]{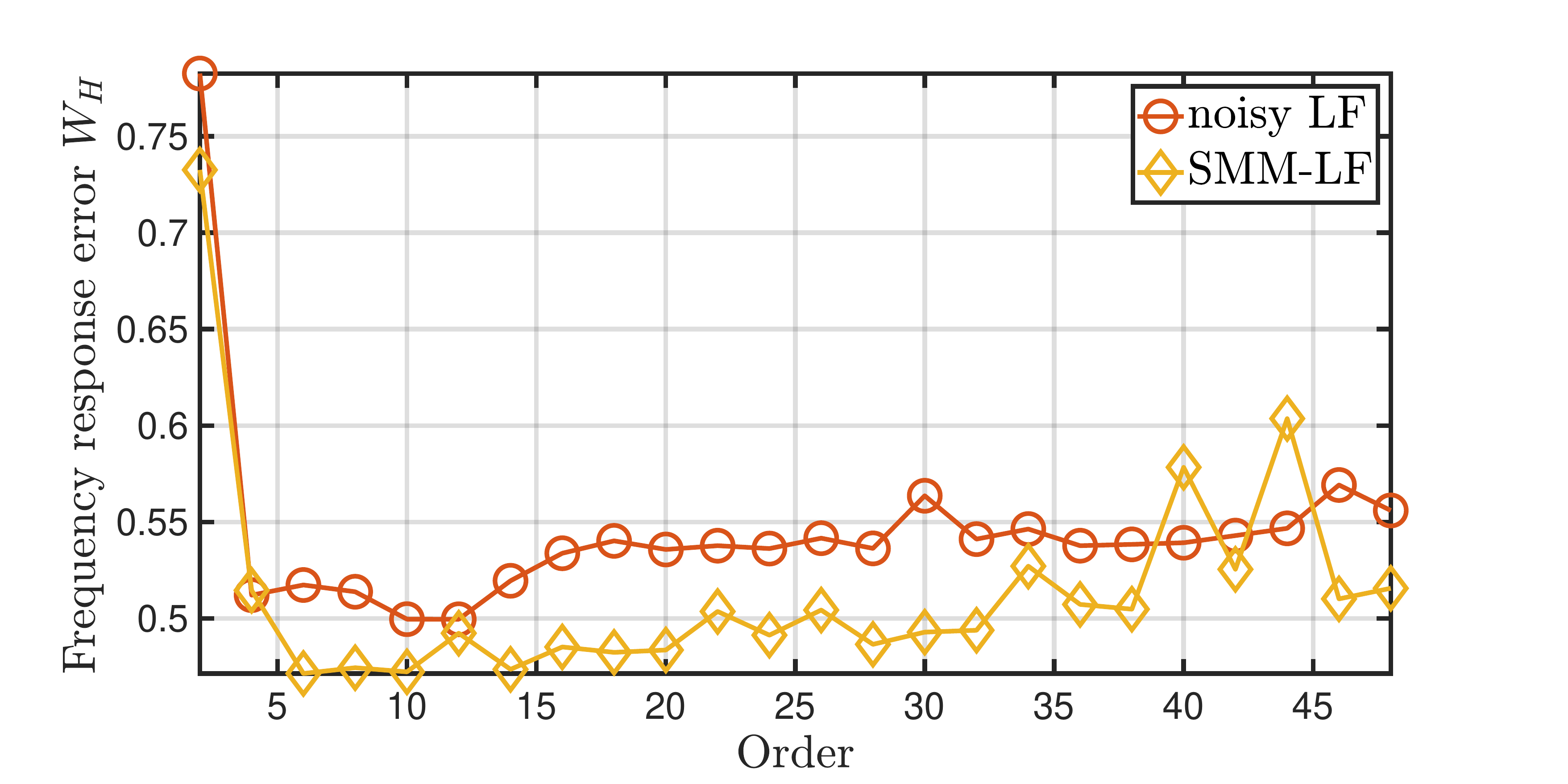}
	\vspace{-3mm}
	\caption{Evolution of the average frequency-domain error $W_H$ according to the reduction order for the SMM-LF and noisy LF procedures.}
	\label{fig:LF_freq_error_order}
	\vspace{-3mm}
\end{figure}

As the LF interpolates the frequency-response, the accuracy of the resulting models is evaluated in the frequency-domain by $W_H$, the normalized $H_2$-error between their frequency-response $\hat{\bH}(e^{\imath\omega_i})$ and the one of the system $\bh(e^{\imath\omega_i})$, evaluated at 200 frequencies log-spaced between 1 and 100rad.s$^{-1}$:
\begin{equation}
\label{eq:H2err_FD}
W_H = \sqrt{\frac{\sum_{i=0}^{N-1} (\hat{\bH}(e^{\imath\omega_i}) - \bH(e^{\imath\omega_i}) )^2}{\sum_{i=0}^{N-1} \bH(e^{\imath\omega_i})^2}}.
\end{equation}
The fitting boxplot is given on Figure \ref{fig:WH_SMM_LF}, to visualize the statistical properties of the error $W_H$ over the 50 noisy collected datasets: the central mark indicates the median, and the bottom and top edges of the box indicate the 25th and 75th percentiles, respectively. The whiskers extend to the most extreme data points not considered outliers, and the outliers are plotted individually. The proposed approach allows to obtain a better fit of the frequency response in average. This is also highlighted by Figure \ref{fig:LF_freqresp} which represents the average frequency response of the resulting models. In addition, the average value of $W_H$ for different reduction order is visible on Figure \ref{fig:LF_freq_error_order}, showing that the proposed approach also leads to more accurate ROMs than noisy LF.

\subsubsection{b) Hankel framework:}
\begin{figure}[h]
	\centering
	\includegraphics[width=0.45\textwidth]{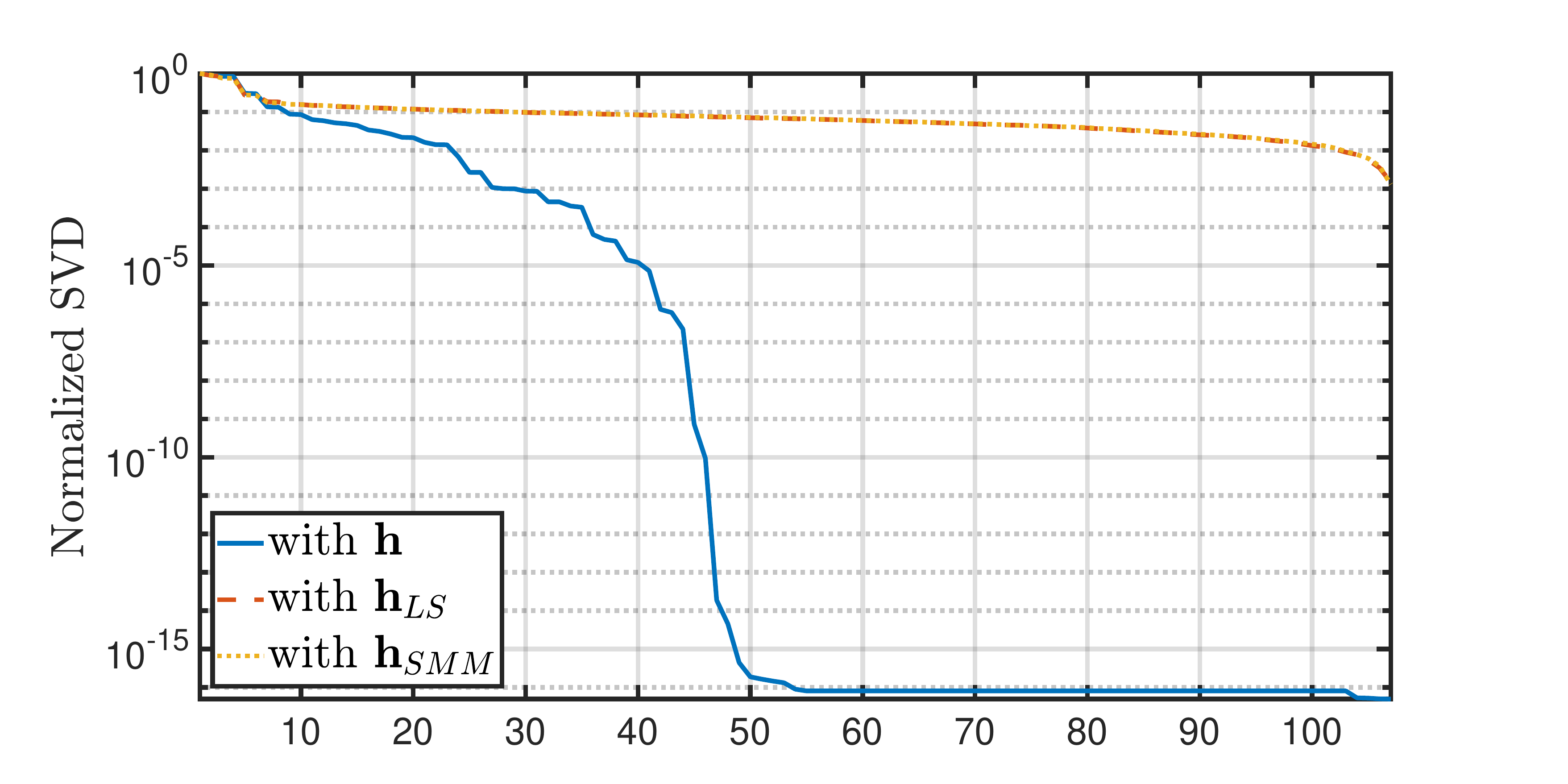}
	\vspace{-3mm}
	\caption{Normalized SVD of the Hankel matrices built with Markov parameters estimated through the LS and SMM approaches and with the true Markov parameters of the system.}
	\label{fig:SVD_hankel}
\end{figure}

The Hankel pencil from \eqref{eq:Hankel_pencil} is built and a SVD is performed on the Hankel matrix $\mathcal{H}$ to reveal the order of the underlying system. The average decay of the normalized singular values is visible on Figure \ref{fig:SVD_hankel} for the true Markov parameters $\bh$ of the system and the estimated ones $\bh_{LS}$ and $\bh_{SMM}$. 

\begin{figure}[h]
	\centering
	\includegraphics[width=0.45\textwidth]{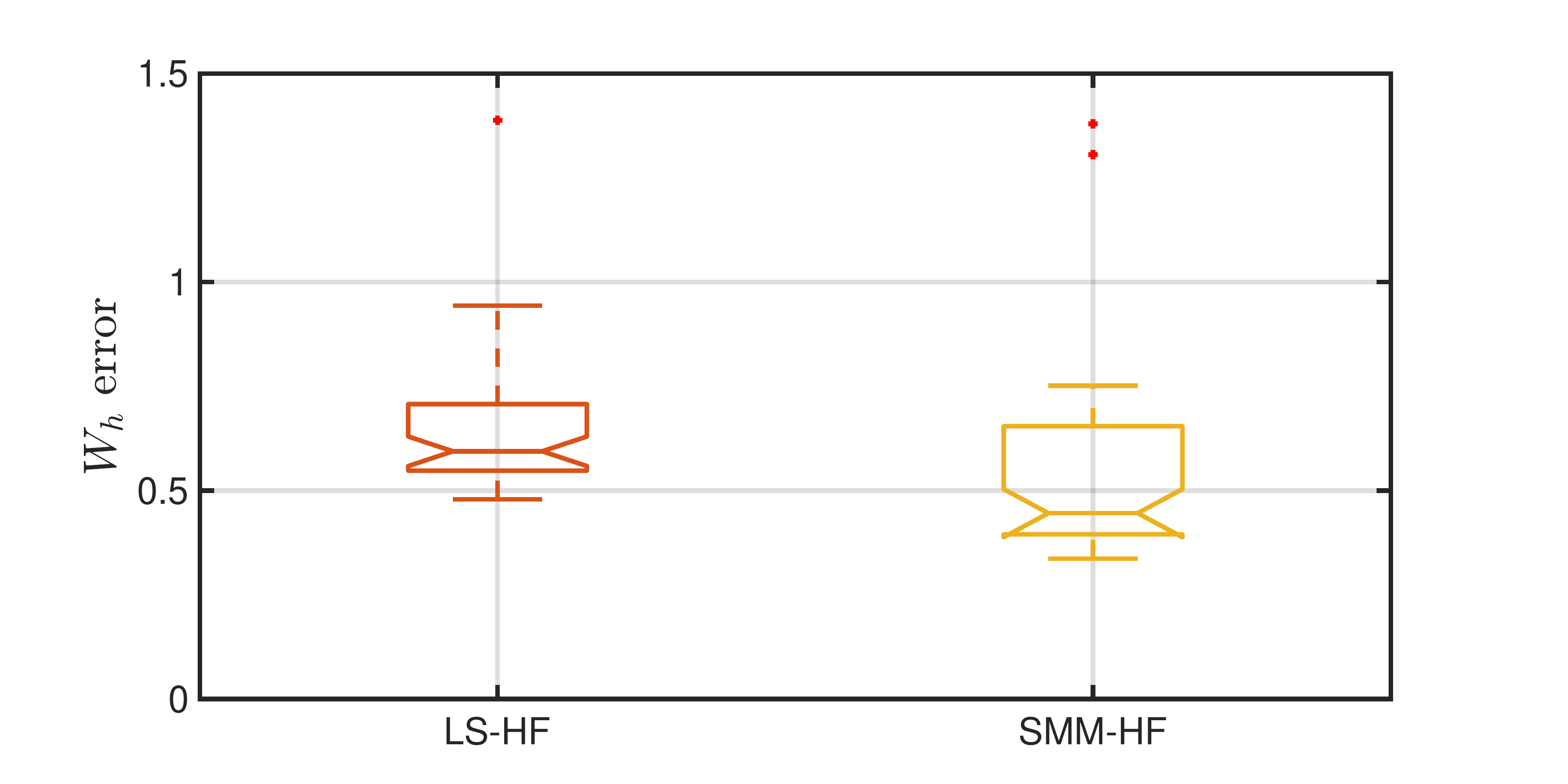}
	\vspace{-3mm}
	\caption{Normalized $\mathcal{H}_2$-error $W_h$ between the resulting impulse responses and the one of the true system.}
	\label{fig:Wh_SMM_HF}
	\vspace{-3mm}
\end{figure}

\begin{figure}[h]
	\centering
	\includegraphics[width=0.45\textwidth]{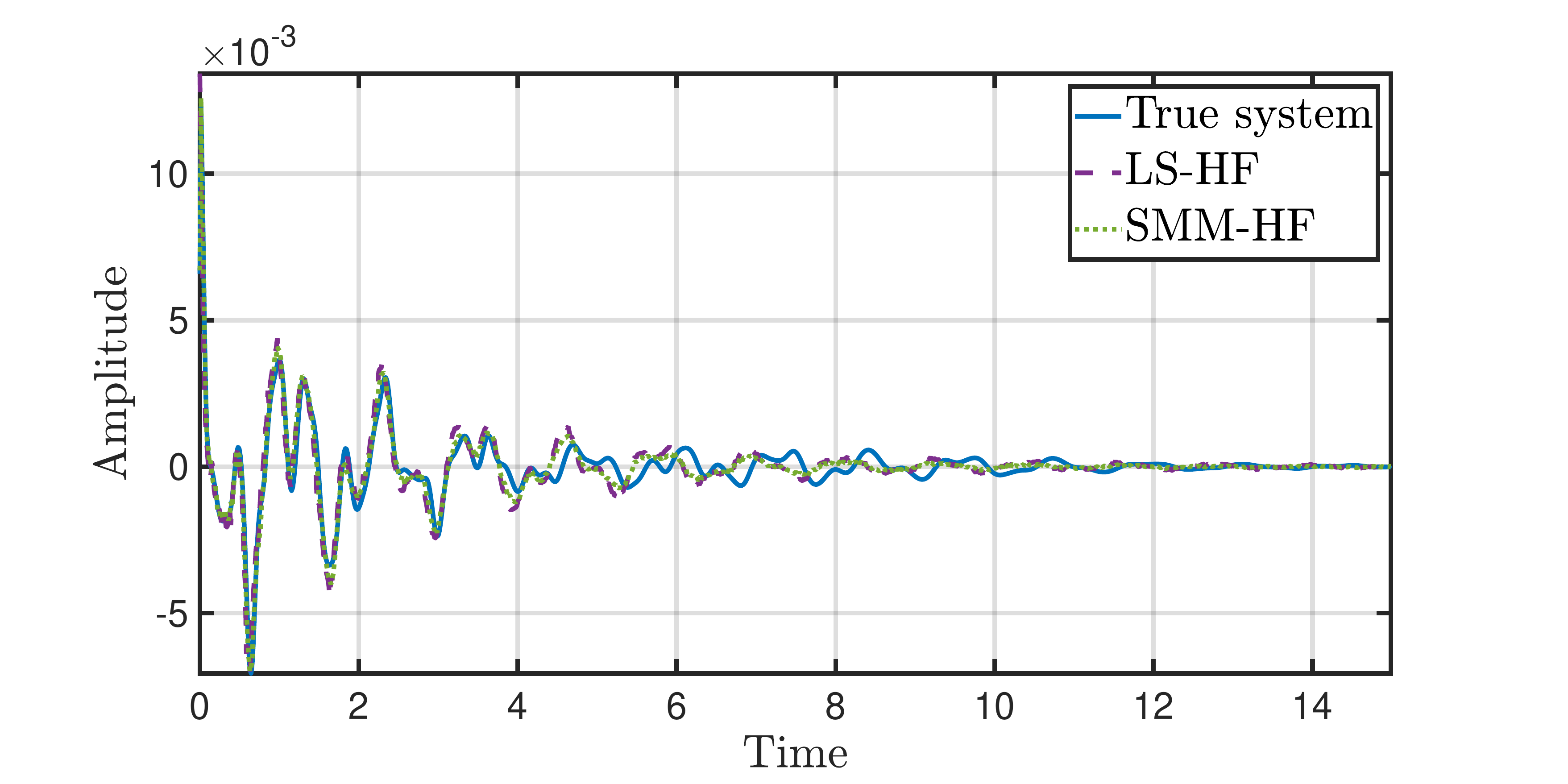}
	\vspace{-3mm}
	\caption{Average impulse response obtained when applying the SMM-HF and LS-HF procedures.}
	\label{fig:HF_impulse}
\end{figure}

\begin{figure}[h]
	\centering
	\includegraphics[width=0.45\textwidth]{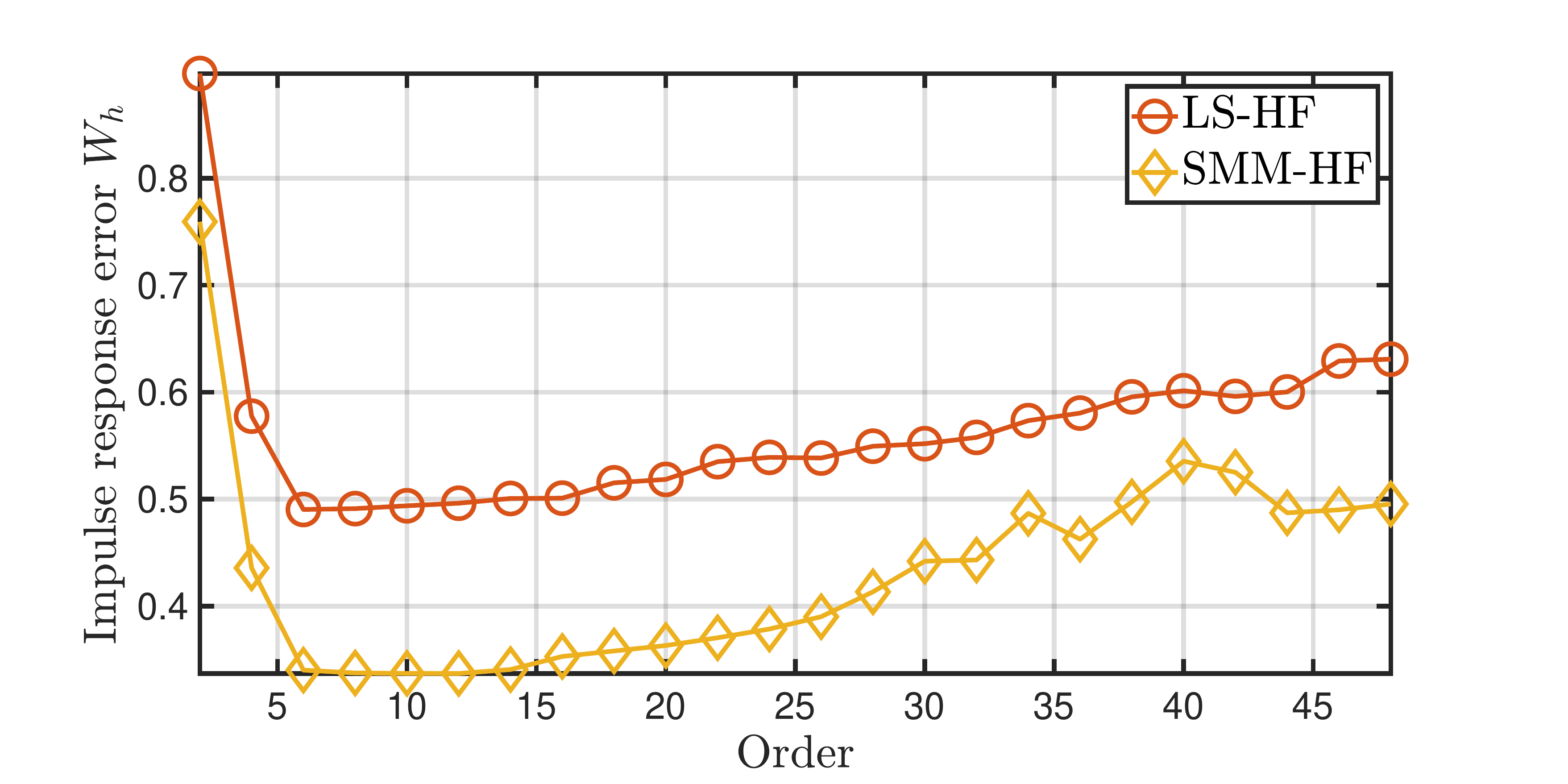}
	\vspace{-3mm}
	\caption{Evolution of the average impulse response error $W_h$ according to the reduction order for the SMM-HF and LS-LF procedures.}
	\label{fig:HF_impulse_error_order}
\end{figure}

The same orders than for LF are chosen. As the HF interpolates the frequency-response, the accuracy of the resulting models is evaluated in time-domain by $W_h$, defined the normalized $H_2$-error between their impulse response $\hat{\bh}$ and the one of the system $\bh$:
\begin{equation}
\label{eq:H2err_TD}
W_h = \sqrt{\frac{\Vert \hat{\bh} - \bh \Vert_2^2}{\Vert \bh \Vert_2^2}},
\end{equation}
The fitting boxplot is given on Figure \ref{fig:Wh_SMM_HF}, showing that the SMM-HF approach from Algorithm \ref{SMM-HF} outperforms the regular LS + HF approach. The average impulse responses are visible on Figure \ref{fig:HF_impulse}. In addition, the average value of $W_h$ for different reduction order is visible on Figure \ref{fig:HF_impulse_error_order}, showing that the proposed approach also leads to more accurate ROMs than LS-HF.

% \subsection{Impact of the noise level}
% \label{subsec:noise_level}

\section{Conlusions and outlooks}
\label{sec:conclusion}
In this work, a method to handle noisy data in matrix pencils frameworks, namely HF and LF, has been proposed. It relies on the SMM approach to estimate the impulse response of the system from a noisy data set. The impulse response constitutes a non-parameterized model of the system, which is then used in the HF or LF to obtain a parameterized model and to reduce it. As in \cite{gosea2021data} \cite{palitta2021efficient}, different data partitioning can be used to reveal the order of the system. As opposed to existing works such as \cite{lefteriu2010modeling}, \cite{sahouli2018iterative} and \cite{kabir2016loewner}, the new method proposes a preliminary step on the available data (the SMM approach), rather than modifying the way of obtaining the model. A thorough comparison between these methods and the proposed approach is left for future work (both in terms of computational complexity and also of accuracy of computed models). Connections to newly-proposed work in \cite{wilber2021data} could also be investigated (this work combines the classical Prony algorithm with the recently-proposed AAA algorithm mentioned in \cite{gosea2021data}).

Future work will also investigate the impact of noise level on the accuracy of the resulting models and it would be interesting to include pseudospectra analysis \cite{embree2019pseudospectra} in the proposed approach. In addition, this work should be illustrated on real-world datasets. The proposed approach could also be used to improve the robustness to noise in the Loewner Data-Driven Control (L-DDC) framework \cite{kergus2018data}, and to introduce a counterpart based on time-domain data relying on the HF the same way L-DDC relies on LF.

% \begin{ack}
% Place acknowledgments here.
% \end{ack}

%\footnotesize
%\bibliography{ifacconf}             % 

\bibliographystyle{spmpsci}
\bibliography{biblio}  

%bib file to produce the bibliography
% with bibtex (preferred)

%\begin{thebibliography}{xx}  % you can also add the bibliography by hand

%\bibitem[Able(1956)]{Abl:56}
%B.C. Able.
%\newblock Nucleic acid content of microscope.
%\newblock \emph{Nature}, 135:\penalty0 7--9, 1956.

%\bibitem[Able et~al.(1954)Able, Tagg, and Rush]{AbTaRu:54}
%B.C. Able, R.A. Tagg, and M.~Rush.
%\newblock Enzyme-catalyzed cellular transanimations.
%\newblock In A.F. Round, editor, \emph{Advances in Enzymology}, volume~2, pages
%  125--247. Academic Press, New York, 3rd edition, 1954.

%\bibitem[Keohane(1958)]{Keo:58}
%R.~Keohane.
%\newblock \emph{Power and Interdependence: World Politics in Transitions}.
%\newblock Little, Brown \& Co., Boston, 1958.

%\bibitem[Powers(1985)]{Pow:85}
%T.~Powers.
%\newblock Is there a way out?
%\newblock \emph{Harpers}, pages 35--47, June 1985.

%\bibitem[Soukhanov(1992)]{Heritage:92}
%A.~H. Soukhanov, editor.
%\newblock \emph{{The American Heritage. Dictionary of the American Language}}.
%\newblock Houghton Mifflin Company, 1992.

%\end{thebibliography}

% \appendix
% \section{A summary of Latin grammar}    % Each appendix must have a short title.
% \section{Some Latin vocabulary}              % Sections and subsections are supported  
% in the appendices.
\end{document}